\documentclass[twocolumn,showpacs,prb]{revtex4}

\def\nbC{{\mathchoice {\setbox0=\hbox{$\displaystyle\rm C$}%
\hbox{\hbox to0pt{\kern0.4\wd0\vrule height0.9\ht0\hss}\box0}}
{\setbox0=\hbox{$\textstyle\rm
C$}\hbox{\hbox to0pt{\kern0.4\wd0\vrule height0.9\ht0\hss}\box0}}
{\setbox0=\hbox{$\scriptstyle\rm
C$}\hbox{\hbox to0pt{\kern0.4\wd0\vrule height0.9\ht0\hss}\box0}}
{\setbox0=\hbox{$\scriptscriptstyle\rm C$}\hbox{\hbox to0pt{\kern0.4\wd0\vrule
height0.9\ht0\hss}\box0}}}}
\def\nbQ{{\mathchoice {\setbox0=\hbox{$\displaystyle\rm
Q$}\hbox{\raise 0.15\ht0\hbox
to0pt{\kern0.4\wd0\vrule height0.8\ht0\hss}\box0}}
{\setbox0=\hbox{$\textstyle\rm Q$}\hbox{\raise
0.15\ht0\hbox to0pt{\kern0.4\wd0\vrule height0.8\ht0\hss}\box0}}
{\setbox0=\hbox{$\scriptstyle\rm
Q$}\hbox{\raise 0.15\ht0\hbox to0pt{\kern0.4\wd0\vrule
height0.7\ht0\hss}\box0}}
{\setbox0=\hbox{$\scriptscriptstyle\rm Q$}\hbox{\raise 0.15\ht0\hbox
to0pt{\kern0.4\wd0\vrule
height0.7\ht0\hss}\box0}}}}
\def\nbT{{\mathchoice {\setbox0=\hbox{$\displaystyle\rm
T$}\hbox{\hbox to0pt{\kern0.3\wd0\vrule
height0.9\ht0\hss}\box0}} {\setbox0=\hbox{$\textstyle\rm
T$}\hbox{\hbox to0pt{\kern0.3\wd0\vrule
height0.9\ht0\hss}\box0}} {\setbox0=\hbox{$\scriptstyle\rm
T$}\hbox{\hbox to0pt{\kern0.3\wd0\vrule
height0.9\ht0\hss}\box0}} {\setbox0=\hbox{$\scriptscriptstyle\rm T$}\hbox{\hbox
to0pt{\kern0.3\wd0\vrule height0.9\ht0\hss}\box0}}}}
\def\nbS{{\mathchoice {\setbox0=\hbox{$\displaystyle     \rm
S$}\hbox{\raise0.5\ht0%
\hbox to0pt{\kern0.35\wd0\vrule height0.45\ht0\hss}\hbox
to0pt{\kern0.55\wd0\vrule
height0.5\ht0\hss}\box0}} {\setbox0=\hbox{$\textstyle        \rm
S$}\hbox{\raise0.5\ht0%
\hbox to0pt{\kern0.35\wd0\vrule height0.45\ht0\hss}\hbox
to0pt{\kern0.55\wd0\vrule
height0.5\ht0\hss}\box0}} {\setbox0=\hbox{$\scriptstyle      \rm
S$}\hbox{\raise0.5\ht0%
\hboxto0pt{\kern0.35\wd0\vrule height0.45\ht0\hss}\raise0.05\ht0%
\hbox to0pt{\kern0.5\wd0\vrule height0.45\ht0\hss}\box0}}
{\setbox0=\hbox{$\scriptscriptstyle\rm
S$}\hbox{\raise0.5\ht0%
\hboxto0pt{\kern0.4\wd0\vrule height0.45\ht0\hss}\raise0.05\ht0%
\hbox to0pt{\kern0.55\wd0\vrule height0.45\ht0\hss}\box0}}}}
\def\nbZ{{\mathchoice {\hbox{$\sf\textstyle Z\kern-0.4em Z$}}
{\hbox{$\sf\textstyle Z\kern-0.4em Z$}}
{\hbox{$\sf\scriptstyle Z\kern-0.3em Z$}}
{\hbox{$\sf\scriptscriptstyle Z\kern-0.2em Z$}}}}
\begin{document}

\title{Randomly dilute Ising model: A nonperturbative approach}
\author{Matthieu Tissier$^1$, Dominique Mouhanna$^1$, Julien
Vidal$^2$, and Bertrand Delamotte$^1$ }
\address{$^1$ Laboratoire de Physique Th\'{e}orique et Hautes
Energies, CNRS UMR 7589, Universit\'{e}
Pierre et Marie Curie Paris 6, 4 place Jussieu, 75252 Paris Cedex 05
France \\ $^2$ Groupe de
Physique des Solides, CNRS UMR 7588, Universit\'{e}s Pierre et Marie
Curie Paris 6 et Denis Diderot
Paris 7, 2 place Jussieu, 75251 Paris Cedex 05 France}

\begin{abstract} {The $N$-vector cubic model relevant, among others,
to the physics of the  randomly
dilute Ising model is analyzed in arbitrary dimension by means of an
exact renormalization-group
equation. This study  provides a unified picture of its critical
physics  between two and four
dimensions. We give the critical exponents for the three-dimensional
randomly dilute Ising  model
which are in good agreement with  experimental and  numerical data.
The relevance of the cubic
anisotropy in the $O(N)$ model is also treated.}
\end{abstract}


\maketitle

Understanding the effects of a weak quenched  disorder on magnetic
systems is one of the great
challenges of statistical physics.  In this context, much attention
has been paid to the Ising model
with  randomly distributed nonmagnetic impurities, i.e. the random
Ising (RI) model  (see \cite{stinchcombe} and \cite{folk} for 
reviews) described by the
following Hamiltonian~:
\begin{equation} H=-J\sum_{\langle i,j\rangle}
\varepsilon_i\varepsilon_j S_i S_j
\label{randomising}
\end{equation} with $J>0$. In Eq.~(\ref{randomising}),  the $S_i$ are
Ising variables and the
$\varepsilon_i$ are quenched, uncorrelated, random variables taking
the value $1$ with probability
$p$ and $0$ with probability $1-p$.  According to the Harris
criterion \cite{harris1}, the
critical behavior of a pure system is altered by the disorder if its
specific-heat exponent $\alpha$
is positive, as it is the case for the three-dimensional Ising model
for which $\alpha\simeq 0.109$,
$\beta\simeq 0.326$, $\gamma\simeq 1.24$, $\nu\simeq 0.630$ and
$\eta\simeq 0.0335$ (see \cite{pelissetto4}).

Experimentally, there now exist several compounds corresponding to a
RI model. They essentially
consist in crystalline difluoride of iron or manganese where FeF$_2$
or MnF$_2$ is replaced by a nonmagnetic material such as ZnF$_2$. 
Experiments performed on
Fe$_p$Zn$_{1-p}$F$_2$ or
Mn$_p$Zn$_{1-p}$F$_2$ with different values of the concentration  $p$
have revealed  scaling laws
with exponents clearly distinct from those of the pure Ising model.
One has (see \cite{folk}), for
instance, for  Fe$_p$Zn$_{1-p}$F$_2$~: $-0.12\le \alpha\le -0.06$,
$0.34\le \beta\le 0.38$, $1.28\le
\gamma\le 1.4$ and $0.68\le \nu\le 0.72$.

Numerically, early Monte Carlo simulations had confirmed the
relevance of disorder  for the Ising
model (see \cite{folk}). The situation was however unclear since the 
critical exponents
seemed to vary with $p$.
Recently, Ballesteros et al. \cite{ballesteros}, by a careful infinite
volume extrapolation of
the data based upon  an analysis of the correction-to-scaling terms,
have reached the conclusion of
dilution-independent critical exponents~: $\alpha=-0.051(16)$,
\mbox{$\beta=0.3546(28)$},
\mbox{$\gamma=1.342(10)$}, $\nu=0.6837(53)$ and $\eta=0.0374(45)$ for
$0.4\le p \le 0.9$.

  From a theoretical point of view, the continuous field theory
relevant to the study of the RI model
is obtained by considering $N$ replica of the model
(\ref{randomising}) and by taking the average
over all possible realizations of the  Gaussian distributed disorder.
This leads to the
Ginzburg-Landau-Wilson action for the $N$-vector model with cubic
anisotropy \cite{grinstein,aharony}~:
\begin{eqnarray} S=\int d^dx \left\{ \displaystyle{1\over 2}
\sum_{a=1}^{N} \left[(\partial_{\mu}
\phi_a)^2+m^2 \phi_a^2\right]+
\right.
\label{randommass}
\\
\left. u_1 \left(\sum_{a=1}^{N} \phi_a^2\right)^2+u_2
\sum_{a=1}^N\phi_a^4\right\}
\nonumber
\end{eqnarray}
with $u_1\propto -(1-p)<0$ and $u_2>0$.  The physics
of the RI model is then obtained
through the  $N\to 0$ limit  of the renormalization group (RG)
equations obtained from  action
(\ref{randommass}). The determination of these RG equations has been
the subject of an intense
activity, mainly performed  within perturbative frameworks.
Different  renormalization schemes have
been used~: minimal subtraction (MS)
scheme \cite{grinstein,khmelnitskii,kleinert,shalaev1,folk4}  or massive
scheme \cite{mayer,pakhnin,pelissetto3,carmona,varnashev1,varnashev2} 
directly in three dimensions where
the $\beta$ functions have been computed up to five \cite{kleinert} 
and  six loops \cite{carmona}
respectively. In each scheme, two different kinds of expansions for
the corresponding $\beta$
functions have been considered~:  {\it i}) $\epsilon$-expansion
(actually
$\sqrt{\epsilon}$-expansion) 
\cite{grinstein,khmelnitskii,kleinert,shalaev1,folk4} in the
MS scheme and pseudo-$\sqrt{\epsilon}$-expansion \cite{folk3} in the 
massive scheme
{\it ii}) expansion in the coupling constants in three
dimensions 
\cite{mayer,pakhnin,pelissetto3,carmona,varnashev1,varnashev2}. It 
has,
however, appeared that the series obtained in case {\it i}) are 
non-Borel summable and
thus  inappropriate for the quantitative evaluation of critical exponents
behavior \cite{folk2,folk4,shalaev1}, a result anticipated by several
authors \cite{mckane,bray} within an analysis of the zero-dimensional 
problem (see
however \cite{alvarez1}).  In case  {\it ii}), genera\-lizations of 
Pad\'e-Borel resummation
techniques for seve\-ral variables have been  considered. In the MS 
scheme, Chisholm-Borel
resummation scheme has been  used \cite{folk1,folk2,folk4}. This 
method has led to
satisfying results at low order but has also appeared to be  unstable 
with respect to the
order of approximation. Indeed, the five-loop calculation does not 
allow to  obtain real
coordinates for the RI fixed point. This point has been related to the possible
Borel-non-summability of the series \cite{folk,folk1,folk2}. In the 
massive scheme,
Pelissetto and Vicari \cite{pelissetto3}, using a procedure of 
resummation formulated within a
subsequent analysis  of the zero-dimensional problem \cite{alvarez1}, 
have employed
Pad\'e-Borel and conformal Pad\'e-Borel resummation techniques 
within a  six-loop RG
calculation. They found a fixed point with  exponents $\gamma=1.330(17)$,
$\nu=0.678(10)$, $\eta=0.030(3)$  that are in rather good agreement
with previous perturbative results \cite{mayer,pakhnin,varnashev1} and with the
present-day numerical data. Let us however mention that, even in this 
case, the location
of  the fixed point is found to be calculation-dependent. This has 
led  Pelissetto et al.
to consider  the possibility of an optimal truncation, possibly 
shorter than that
considered in \cite{pelissetto3}, a fact that could
constitute a limitation of the method.

Despite the good agreement between experimental and numerical data
and the  perturbative results, we
would like to stress upon the need for  a complementary approach, not
relying on perturbative
expansions and this, for several reasons. The first one is that the
perturbative field theoretical
treatment implies the  use of sophisticated resummation techniques
which, if they proved to lead to
safe results in the case of pure magnets, are far from being under
control in the case of disordered
systems \cite{bray,mckane,alvarez1,folk}. The second reason for the use of a
nonperturbative treatment  is that there exist cases where
the perturbative approach apparently seems to provide incorrect
results. For instance, in the context
of frustrated magnets, six-loop perturbative RG computations predict
the existence of a fixed
point \cite{pelissetto1} and, thus, a universal behavior for their
critical physics. The
experimental data, although displaying scaling laws, are rather in
favor of a  nonuniversal
behavior, the critical exponents varying from one material to the
other.  Computations based on the
exact RG formalism have shown to be able to reproduce this
phenomenology \cite{tissier2,tissier3}. Finally, there are situations
where only  specific
techniques  work well. For example, in two dimensions, one can use 
integrability or
conformal invariance which, however, only operate in this dimension.
All these features  plead toward
the use of a method that does not rely on the treatment of divergent
series, provides a
systematic way to improve the results and is able to tackle with
the physics in any dimension.

A Wilson-like \cite{wilson} nonperturbative treatment of
(\ref{randommass}) has been first
performed by Newman and Riedel \cite{newman} on the basis on the
scaling-field (SF) method.
However, due to the complexity of its practical implementation, this
approach has been given up and
perturbative approaches largely preferred. We used here a
nonperturbative approach based on the
concept of effective average action
$\Gamma_k[\phi]$  \cite{wetterich2,berges3} which appears as a very
efficient tool. The  quantity
$\Gamma_k[\phi]$ has the meaning of a  coarse-grained free energy at the scale
$k$ in the sense that it only  includes the fluctuations with momenta
$q\ge k$. It is a function of
the effective order parameter  $\phi$ at the scale $k$, the analog of
a magnetization at this scale.
In the low-energy limit $k\to 0$, all modes are integrated out so
that $\Gamma_{k=0}$ identifies with
the usual effective action $\Gamma$. At the microscopical scale
$\Lambda$, no fluctuation has been
integrated and $\Gamma_{k=\Lambda}$ identifies with the microscopic
action $S$. The $k$-dependence of
$\Gamma_k$ is controlled by an exact evolution equation \cite{wetterich1}~:
\begin{equation} {\partial \Gamma_k\over \partial t}={1\over 2} \hbox{Tr}
\left\{(\Gamma_k^{(2)}+R_k)^{-1}
   {\partial R_k\over \partial t}\right\}\ ,
\label{renorm}
\end{equation} where $t=\ln \displaystyle ({k / \Lambda})$. The trace
has to be understood as a momentum integral as well as a summation 
over internal indices. In
Eq.(\ref{renorm}), $\Gamma_k^{(2)}$, the
second derivative of $\Gamma_k$, is  the {\it exact field-dependent}
inverse propagator and $R_k$ is the running infrared cutoff that 
suppresses the propagation of
modes with momenta $q<k$. A
convenient cutoff is provided by \cite{berges3}~: $R_k(q)=Z 
q^2/[\exp(q^2/k^2)-1]$, where $Z$ is
the field renormalization. The effective average action $\Gamma_k$ is
a functional invariant under
the symmetry group of the system and thus includes all powers of all
invariants -- and their
derivatives -- built out from the average order parameter. Thus,
Eq.(\ref{renorm}) is a nonlinear
functional  equation, too difficult to be solved exactly so that, in
practice, one has to deal with
a truncated form of $\Gamma_k$. We choose here to perform  an
expansion of $\Gamma_k$ around  its
minimum, keeping  a finite number of monomials in the invariants and
including the most relevant
derivative terms \cite{berges3}.  We first consider the following truncation~:
\begin{eqnarray}
\Gamma_k&=& \displaystyle \int d^dx \left\{\sum_{a=1}^N {Z\over 2}
(\partial_{\mu}{\phi_{a}})^2+u_{1}\left[\sum_{a=1}^{N}
({\phi_{a}}^2-\kappa)\right]^2 \nonumber
\right.
\\ &&+ \left. u_{2}\sum_{a=1}^{N}\left({\phi_{a}}^2-\kappa\right)^2\right\}
\label{action}
\end{eqnarray} where $\left\{u_{1},u_{2},\kappa,Z\right\}$  are the
$k$-dependent coupling constants
that parametrize the  evolution of the model with  the scale $k$. The
action (\ref{action}) displays
a discrete ${\nbZ}_N$ permutation symmetry. For $N u_1+u_2 >0$, the
minimum  of action (\ref{action})
is given by a replica symmetric configuration defined by
$\phi_{a}^{min}=\sqrt{\kappa}$ for all $a$.
Expression (\ref{action}) thus corresponds to a  quartic expansion of
$\Gamma_k$ around the minimum
$\phi_{a}^{min}$.  There is no residual symmetry in the minimum so
that the symmetry breaking scheme
is given by a fully broken ${\nbZ}_N$ group. The spectrum of the
theory in the broken phase is given
by the eigenvalues of $\Gamma_k^{(2)}$ at the minimum and consists in
a singlet of mass
$m_1^2=8\kappa (N u_{1}+u_{2})$ and a $(N-1)$ uplet of mass
$m_2^2=8\kappa  u_{2}$. When $u_2=0$,
this $(N-1)$ uplet cor\-responds to the Goldstone modes of the
$O(N)$-invariant model that gets here
a mass proportionnal to the $O(N)$-symmetry-breaking-coupling-constant $u_2$.

Using Eq.(\ref{renorm}) and Eq.(\ref{action}), we have derived the
flow equations for the
dimensionless renormalized coupling constants $\kappa_r$, $u_{1r}$,
$u_{2r}$, and  $Z$~:
\begin{widetext}
\begin{eqnarray}
\partial_t \kappa_r &=& -(d - 2 + \eta)\kappa_r - {2 v_d\over N(N
u_{1r}+u_{2r})} \bigg[3(N
u_{1r}+u_{2r}) L_1^d[m_{1r}^2]+(N-1)(N u_{1r}+3 u_{2r})
L_1^d[m_{2r}^2]\bigg]\nonumber \\
\partial_t u_{1r}&=&(d-4+2 \eta) u_{1r}-{2 v_d\over N^2
\kappa_r}\bigg[- N u_{1r} L_1^d[m_{1r}^2]+ N
u_{1r} L_1^d[m_{2r}^2]+36 \kappa_r  (N u_{1r}+u_{2r})^2
L_2^d[m_{1r}^2] \nonumber \\ & &+\bigg(4(
N-1) N^2 \kappa_r  u_{1r}^2+24 N(N-1)\kappa_r  u_{1r} u_{2r}+36
\kappa_r u_{2r}^2\bigg)
L_2^d[m_{2r}^2]- 8 \kappa_r (N u_{1r}+3 u_{2r})^2
L_{1,1}^d[m_{1r}^2,m_{2r}^2]\bigg]
\label{recursion}\\
\partial_t u_{2r}&=&(d-4+2 \eta) u_{2r}-{2 v_d\over N
\kappa_r}\bigg[N u_{1r} L_1^d[m_{1r}^2] -N
u_{1r} L_1^d[m_{2r}^2]+36 (N-2) \kappa_r  u_{2r}^2 L_2^d[m_{2r}^2]
\nonumber\\ & &+8 \kappa_r (N
u_{1r}+3 u_{2r})^2 L_{1,1}^d[m_{1r}^2,m_{2r}^2]\bigg]\nonumber\\
\eta&=&-\partial_t \ln Z=-{256 \kappa_r  v_d\over N d}\bigg[(N-1)
u_{2r}^2 M_{2,2}^d[0,m_{2r}^2]+ (N
u_{1r} +u_{2r})^2 M_{2,2}^d[0,m_{1r}^2]\bigg]\nonumber
\end{eqnarray}
\end{widetext} where $1/v_d=2^{d+1} \pi^{d/2} \Gamma(d/2)$. In
Eqs.(\ref{recursion}), $L_{n_1,
n_2}^d[m_{1r}^2,m_{2r}^2]$, $L_{n}^d[m_{ir}^2]$  and 
$M_{n_1,n_2}^d[m_{1r}^2,m_{2r}^2]$ are the
``threshold  functions'' \cite{tetradis3} that
encode the nonperturbative aspects of the RG equations
(\ref{recursion}). We show how this set of equations provides a
qualitative and, partly, quantitative
description of the physics of model  (\ref{randommass}) between $d=2$
and $d=4$.

Around $d=4$, the fixed point coupling constants $u_{1r}$ and
$u_{2r}$ are expected to be of order
$\epsilon=4-d$. One can thus obtain simplifications of
Eqs.(\ref{recursion}) performing a small mass
-- $m_{1r}$ and $m_{2r}$ -- expansion of the threshold functions.
Using their asymptotic behavior one
finds \mbox{$\eta=O(u_{1r}^2,u_{2r}^2)$} and~:
\begin{eqnarray} {\partial}_t u_{1r}&=&-\epsilon u_{1r}+{1\over 2
\pi^2}\left[(N+8)u_{1r}^2 +6
u_{1r}u_{2r}\right]
\label{rg1}
\\
\nonumber\\ {\partial}_t u_{2r}&=&-\epsilon u_{2r} + {1\over 2
\pi^2}\left[9 u_{2r}^2+12 u_{1r}
u_{2r}\right]\ .
\label{rg2}
\label{4d}
\end{eqnarray} These equations are those obtained directly in the
weak-coupling expansion of the
$N$-vector cubic model \cite{grinstein,aharony}. For $N\ne 0$ they
display three nontrivial
fixed points~: the $O(N)$ fixed point $(u_{1r}^*>0,u_{2r}^*=0)$, the
$N$-decoupled Ising one
$(u_{1r}^*=0,u_{2r}^*>0)$ and the cubic one $(u_{1r}^*>0,u_{2r}^* 
\neq 0)$. The cubic fixed
point is stable for $N>N_c$ and unstable for  $N<N_c$ with
$N_c(d=4-\epsilon)=4-2\epsilon+O(\epsilon^2)$, as  known from early
perturbative
approaches \cite{grinstein,aharony}. For $N=0$ the only fixed point of
the equations (\ref{rg1})
and (\ref{rg2}) is the Gaussian one but this is an artefact of the
leading order of perturbation
theory that disappears at second order in $\epsilon$. This leads to the famous
$\sqrt\epsilon$-expansion \cite{khmelnitskii,grinstein}. However, when the
Eqs.(\ref{recursion}), obtained
with the simplest ansatz (\ref{action}), are analyzed without
$\epsilon$-expansion, the random fixed
point already appears. All the fixed points found here persist when
the dimension is lowered from
$d=4$ to the vicinity of $d=2$, a case that deserves a special 
treatment \cite{shalaev}.

Around $d=2$, one also expects the Eqs.(\ref{recursion}) to well
describe, even quantitatively, the
physics in the vici\-nity of the $O(N>1)$ fixed-point. The  reason is
that, around this dimension, the
phase transition associated to this fixed point takes place at a
temperature of order $\epsilon=d-2$.
Since the anomalous dimension is of the same order of magnitude, this
situation must be well
described even with our wild truncation that neglects high order
derivative terms. Around $d=2$,
$1/(N\kappa_r)= T_r$ provides the effective perturbative temperature
of the associated $O(N)$
non-linear sigma (NL$\sigma$)  model. One also has to consider $u_2$
as a small perturbation around
the $O(N)$ fixed point. Thus, the procedure now consists in a double
expansion  of
Eqs.(\ref{recursion}) in the large mass $m_{1r}$ and in the small
coupling constant $u_{2r}$. Using
the asymptotic behaviors of the threshold functions, one obtains~:
$\eta={T_r/(2\pi N)}$ and~:
\begin{eqnarray} {\partial}_t T_r&=&\epsilon T_r-{(N-2)\over 2 \pi} T_r^2
\\ {\partial}_t u_{1r}&=&-2 u_{1r} +4 u_{1r}^2 {(N-1)\over \pi}\ln2
\\ {\partial}_t u_{2r}&=&\left(-2 + \epsilon  + {4\over N \pi} T_r
\right) u_{2r}\ .
\label{2d}
\end{eqnarray}

These equations admit a nontrivial fixed point with
$T_r^*=2\pi \epsilon/(N-2)$,
$u_{1r}^*=\pi/(2(N-1) \ln 2 )$ and $u_{2r}^*=0$ which corresponds to
the standard $O(N)$ fixed point
of the NL$\sigma$ model. It is then interesting to consider the flow
of $u_{2r}$ around this fixed
point. One has~:
\begin{equation} {\partial}_t \delta
u_{2r}=\left(-2+{(N+6)\over(N-2)}\ \epsilon\right) \delta
u_{2r}\ .
\end{equation}  This shows that the cubic perturbation is relevant
for $N>N_c$ and irrelevant for
$N<N_c$ with $N_c(d=2+\epsilon)=2+4\epsilon+O(\epsilon^2)$, in
agreement with the perturbative
approach of the NL$\sigma$ model around $d=2$ \cite{pelcovits}.
Concerning the fixed points with
discrete symmetry (Ising $O(1)$, $N$-decoupled Ising, cubic and
random Ising), they cannot, around
$d=2$, be properly described  by our truncation. Actually, the large
values of $\eta$ at these fixed
points around $d=2$ likely require the inclusion  in (\ref{action})
of all powers of the field and
high order derivative terms.

In $d=3$, in order to obtain satisfying quantitative results, we have
pushed the calculation using a
more sophisticated ansatz than
   (\ref{action}).  We have considered all monomials up to order eight
in the fields  as well as all derivative terms including two 
derivatives and four
fields and established
   the RG equations for the corresponding coupling constants. These
equations are too long
   to be given here \cite{tissier4} and we just give our results. First, we have
   tested our calculation on the $O(N)$ model. In the Ising and
Heisenberg cases, we have found $\nu=0.627$ and $\nu=0.719$, 
respectively, which are very
reasonable compared to
   previous results(see  \cite{pelissetto4}). The values of the anomalous
dimensions $\eta=0.053$ and
$\eta=0.0452$ are less convin\-cing and related to the omission of higher
   order derivative terms in $\Gamma_k$.  Considering the cubic
anisotropy, we have determined an
estimate of $N_c$ above which the cubic fixed point is stable and
found $N_c(d=3)=3.1$. This value
differs significantly from those obtained from recent perturbative 
approaches for which
$N_c<3$ 
\cite{folk3,mayer,carmona,varnashev1,varnashev2,shalaev1,kleinert}. 
For instance,
in a recent six-loop calculation, $N_c$ is found to be equal to 
2.89(4) \cite{carmona}, a
value already obtained from lower order computations 
\cite{varnashev1,varnashev2}. If
this last result is confirmed, our result should be imputed to the 
poor determination of
$\eta$. The strong dependence of $N_c$ with respect to $\eta$ already 
appears in a
hierarchical approximation where,
neglecting any field renormalization, $N_c=2.219$ \cite{pinn}.
   Note also that the result $N_c=3.4$ was found within the work
   of Newman and Riedel \cite{newman}. For the RI
   fixed point we have found $\nu=0.67$ and $\eta=0.05$ that give,
through scaling laws,
   $\gamma=1.306$, $\beta=0.351$ and $\alpha=-0.01$. These results are
in good agreement with perturbative
methods 
\cite{folk,folk2,folk4,pelissetto3,varnashev1,varnashev2,pakhnin,mayer 
} and also
with numerical estimates, while  they show a certain discrepancy with 
the SF method that
provides $\nu=0.697$, $\gamma=1.39$ and $\eta=0.015$ \cite{newman}.

Finally, we indicate that we have noted a rapid convergence of our 
results with respect to the order of
the truncation in the fields. However, a challenging and still open 
question is that of the
convergence when including higher order derivative terms which is 
essential for an improvement of the
anomalous dimension. \\

We thank M. L. Rosinberg, B. N. Shalaev and  G. Tarjus for discussions.

\end{document}